# A Computed Tomography Vertebral Segmentation Dataset with Anatomical Variations and Multi-Vendor Scanner Data

Hans Liebl[1*], David Schinz[1*], Anjany Sekuboyina[1, 2], Luca Malagutti[1], Maximilian T. Löffler[3], Amirhossein Bayat[1,2], Malek El Husseini[1,2], Giles Tetteh[1, 2], Katharina Grau[1], Eva Niederreiter[1], Thomas Baum[1], Benedikt Wiestler[1], Bjoern Menze[2], Rickmer Braren[4], Claus Zimmer[1], Jan S. Kirschke[1]

[1] Department of Diagnostic and Interventional Neuroradiology, School of Medicine, Klinikum rechts der Isar, Technical University of Munich, Germany

[2] Department of Informatics, Technical University of Munich, Germany

[3] Department of Diagnostic and Interventional Radiology, University Medical Center Freiburg, Freiburg im Breisgau, Germany

[4] Department of Diagnostic and Interventional Radiology, School of Medicine, Klinikum rechts der Isar, Technical University of Munich, Germany

[*]both authors equally contributed to this manuscript

## Abstract:

With the advent of deep learning algorithms, fully automated radiological image analysis is within reach. In spine imaging, several atlas- and shape-based as well as deep learning segmentation algorithms have been proposed, allowing for subsequent automated analysis of morphology and pathology. The first "Large Scale Vertebrae Segmentation Challenge" (VerSe 2019) showed that these perform well on normal anatomy, but fail in variants not frequently present in the training dataset. Building on that experience, we report on the largely increased VerSe 2020 dataset and results from the second iteration of the VerSe challenge (MICCAI 2020, Lima, Peru).

VerSe 2020 comprises annotated spine computed tomography (CT) images from 300 subjects with 4142 fully visualized and annotated vertebrae, collected across multiple centres from four different scanner manufacturers, enriched with cases that exhibit anatomical variants such as enumeration abnormalities (n=77) and transitional vertebrae (n=161). Metadata includes vertebral labelling information, voxel-level segmentation masks obtained with a human-machine hybrid algorithm and anatomical ratings, to enable the development and benchmarking of robust and accurate segmentation algorithms.

## Abbreviations:

BMD= bone mineral density, CADx= computer-aided diagnosis, CNN= convolutional neural network, CT= computed tomography, GCN= graph convolutional network, MICCAI= Medical Image Computing and Computer Asisted Intervetion, PC= point cloud analysis

# Background and Summary

Numerous applications of computer-aided diagnostics (CADx) are currently being developed beginning to gradually reshape the future of radiological clinical practice and research.[1-7] In spinal imaging, different deep learning approaches have been used for vertebral labelling and segmentation tasks in the form of convolutional neural networks (CNN), graph convolutional networks (GCN) or point clouds (PC) to analyse bone structures.[8-13] Various applications of computer algorithms have shown great potential to detect vertebral fractures and to measure bone mineral density (BMD).[13-17] However, most of these approaches are largely data dependent, as the algorithms require extensive datasets with corresponding metadata for the development, training and validation to enable efficient models.[18] Aiming at the task of improving automated quantification of spinal morphology and pathology by vertebral labelling and segmentation, the first iteration of the "Large Scale Vertebrae Segmentation Challenge" (VerSe 2019)[17,19] was held at the International Conference on Medical Image Computing and Computer Assisted Intervention (MICCAI 2019, Shenzhen, China). The segmentation challenge received considerable participation from the scientific community with more than 250 registrations and 20 participating teams.[19] As part of the VerSe 2019 challenge a large dataset was provided addressing the previous severe shortage of publicly-available, large, accurately annotated CT spine data in the community by releasing 160 CT image series and their voxel-level annotations comprised of a large variety in fields of view and spatial resolutions as well as spinal and vertebral pathologies.[20]

Building on the data, experience, and learning from the VerSe 2019 challenge, we proposed to organise a second iteration of the vertebrae segmentation challenge at the MICCAI 2020 in Lima, Peru. For the VerSe 2020 challenge, we aimed to substantially increase the existing dataset to 300 subjects. While retaining the richness of the VerSe 2019 dataset, the VerSe 2020 dataset was complemented with images from different institutions and four major scanner manufacturers. In addition, we aimed to include rare anatomical variants such as numeric aberrations and cervicothoracic or lumbosacral transitional vertebrae. As such variants have a low prevalence in the population, they are under-represented, if present at all, in unselected training datasets like that presented at the VerSe 2019 challenge. Consequently, deep learning-based algorithms fail in such cases. Thus, focus was given to enrich the VerSe 2020 dataset with rare anatomical variants to improve derived model performance as previously described.[21] We envisioned the creation of an extended annotated CT dataset to provide consistent and reliable ground truth data for algorithm training and benchmarking.

The proposed Verse 2020 dataset was released as part of the second iteration of the "Large Scale Vertebrae Segmentation Challenge" hosted at the MICCAI conference held in Lima, Peru (https://verse2020.grand-challenge.org/).[19] The dataset was split into a training dataset, a public test dataset, and a private test dataset building on the preexisting VerSe 2019 dataset[20] published for the MICCAI conference in 2019, with an overlap of 105 CT image series now comprising 319 image series of 300 subjects. To date, this dataset represents the largest publicly available CT imaging dataset of the spine with corresponding metadata including labelling information, voxel-level segmentations of all fully visualized vertebrae and definition of enumeration abnormalities and transitional vertebrae.

In summary, the successful segmentation challenges held at the MICCAI conferences in 2019 and 2020 based on these public datasets confirm, that reliable, fully-automated deep learning algorithms for segmentation of the spine can be trained and that algorithm performance benefits from large and diverse datasets.

Therefore, we regard the VerSe 2020 dataset as an important step towards clinical translation of CADx algorithms for spine imaging, which may soon supplement the radiologist's work in daily routine. We are convinced that in the near future, patients will greatly benefit from CADx extracting even more relevant information from medical imaging than currently possible.

# Methods

## *Subject Selection*

This retrospective evaluation of imaging data was approved by the local institutional review board and written informed consent was waived (Proposal 27/19 S-SR).

Inclusion criteria for the dataset: Subjects older than 18 years were included, who had received CT imaging of the spine showing a minimum of 7 fully visualized vertebrae without counting sacral vertebrae or transitional vertebrae. Exam dates were limited to the time between February 5th, 2016 and March 1st, 2020. The minimum required spatial resolution was defined as 1.5 mm in the craniocaudal direction, 1mm in the anterior-posterior direction and 3mm in the left-right direction to allow for a sufficient delineation of vertebral deformities.[22] As in VerSe 2019, traumatic fractures and bony metastases were excluded. Other osseous changes such as Schmorl nodes, hemangioma, degenerative changes, or the presence of foreign materials for kyphoplasty or spondylodesis intentionally remained part of the dataset to reflect the widest possible spectrum of spine morphology. Aiming at a >100% increase compared to the 141 subjects from Verse 2019 and providing a dataset with 50% multivendor data and 50% anatomical variants, we composed a dataset consisting of 300 subjects, including 86 subjects from VerSe 2019 and 214 new subjects. In order to select new subjects, we searched the institutional picture archive and communication system (PACS) regarding two aspects: 1) CT studies imported from other institutions, that were acquired on scanner hardware different from that installed in our institution; 2) CT reports documenting the presence of spinal anatomical variants including enumeration abnormalities, transitional vertebrae or cervical ribs. In both queries, we aimed for a balanced composition of cases: From the first query, we selected 20 subjects from external Toshiba scanners, 20 subjects from external GE scanners and 30 subjects from external Siemens scanners. We added to 30 subjects from Ben Glocker´s [23] dataset and 50 subjects from VerSe 2019 (10 from Siemens scanners and 40 from Philips scanners), to form the multivendor dataset with 150 subjects. The query for anatomical variants of spinal anatomy such as thoracolumbar and lumbosacral transitional vertebrae, cervical ribs, thoracic and lumbar short ribs as well as enumeration variants revealed 308 subjects. Adding to 36 VerSe 2019 subjects with anatomical variants we selected another 114 subjects to form the

150 subjects of the dataset with anatomical variants. The final dataset comprised 300 subjects with a total of 319 image series, as some subjects comprised two separate image series (e.g. thoracic spine and lumbar spine). Subject characteristics and data subset stratification are listed in Table 1. All selected imaging series were categorized regarding their primary attribute (e.g. Toshiba scanner, Castellvi grade 4 transitional vertebra, or numeric aberration with 4 lumbar vertebrae) and each subgroup was randomly split to the training, public validation and private test subsets as demonstrated in Figure 1.

Despite the wide inclusion criteria, very rare variants might still be missing and osseous changes other than the included anatomical variants or cases with new foreign material may still limit its generalizability. Also, the sacrum including lumbar transitional vertebrae, i.e. vertebrae partially fused with the sacrum, have not been segmented in this dataset. Future works, potentially based on the presented data, will need to address complete segmentation of the spine including the sacrum. Another future focus may be the inclusion of more pathological changes, e.g. only two cases with half-vertebrae are included in VerSe 2020, metastases and dislocated traumatic fractures were excluded. In this regard, it first remains to be discussed how these pathologies should be labelled and segmented.

## *Vertebral Segmentation*

The labelling of the vertebrae and the segmentation process are essential steps in processing spine data. All subsequent analyses such as the detection and grading of fractures, calculation of bone mineral density but also analysis of spinal shape, curvature, and deformity such as scoliosis rely on these initial tasks.

For the proposed dataset, we used a semi-automated in-house developed algorithm to generate segmentation masks of the vertebrae step-by-step as illustrated in Figure 2: first, the CT input data was anonymized by conversion to Neuroimaging Informatics Technology Initiative (NIfTI) format (https://nifti.nimh.nih.gov/nifti-1). To ensure full anonymity, defacing was achieved by deleting the raw data in a manually segmented mask of the face. In case of the 86 re-used VerSe 2019 subject datasets, the spatial resolution has previously been reduced to 1 mm isotropic or in sagittal 2 mm/ 3 mm series to 1 mm in-plane resolution. In case of the 214 newly added patient datasets, the original resolution was kept. Second, a deep-learning framework (publicly accessible under: https://anduin.bonescreen.de) was used to label and segment individual vertebrae. In brief, at first a low-spatial resolution CNN is used for the detection of all osseous spinous structures, resulting in a low-resolution heatmap. This was used to automatically generate a spine bounding box containing the spine. Second, the Btrfly Net was used to label the vertebrae, with the option to manually correct the centroids if needed. Third, a U-Net was used in order to segment each vertebra defined by each single vertebral label in separate patches, defined by bounding boxes around the vertebral labels.[11,24-26] These vertebral masks were created at $1mm^3$ isotropic resolution and subsequently merged into one multi-label segmentation mask with individually labelled vertebrae. Originally, both Btrfly Net and U-Net were trained with the VerSe 2019 data and have continuously been improved using the labels and segmentation masks derived from this dataset for repetitive training. Third, the segmentation output of the algorithm was manually corrected in a laborious process by two specifically trained medical students using the open-source software ITK-SNAP[27]. This manual correction was performed in the original imaging space, but was limited to an accuracy of 1mm, similar to the output of the U-net; i.e. segmentation errors in smaller structures than 1mm were not corrected. Finally, corrections were reviewed and corrected or approved by a neuroradiologist to achieve the highest possible consistency of the presented segmentation masks. Despite the good performance of the baseline algorithm, the correction of on average 15 objects of interest with approximately $10^5$ voxels took considerable effort.

## *Anatomical Variants*

As anatomical variants of the spine can be frequently observed, both numeric and morphologic changes have been intentionally included in the VerSe 2020 dataset. Tins and Balain have reported numeric anatomical variations of the spine to be more frequent (7.7%) than transitional vertebrae (3.3%) with a tendency of male subjects to show more additional vertebrae and females to show more missing vertebrae.[28] Numeric aberrations of the cervical spine are rare, but additional cervical ribs can be frequently observed in a clinical setting with a prevalence of 0.05% - 6.1%.[29] According to the literature, variations of the thoracolumbar region are generally rarely reported and often overlooked, despite potential clinical implications e.g. for surgical planning. Variants of the lumbosacral region are frequently observed and according to Thawait et al.[29] can be found in 4%-30% of examined cases. Wigh et al. differentiated the presence of accessory ossification centres and stump ribs as compared to the last pair of ribs at a thoracic vertebra based on length and morphology and defined any vertebra with ribs shorter than 38 mm as transitional thoracolumbar vertebra.[30] While there are different approaches to classify transitional vertebrae, computer algorithms need a clear definition to which part of the spine a vertebra belongs. Therefore, in our dataset vertebrae with ribs larger than 38 mm on either side and a typical diagonal downward alignment were classified as thoracic vertebrae. In cases with both ribs smaller than 38 mm, e.g., when only small ossification centres were present with a horizontal alignment, the vertebra was considered lumbar. In ambiguous cases, additional morphological features were used to identify a vertebra as thoracic or lumbar such as shape features of the vertebra and the orientation of the articular joint facets.

Regarding the lumbosacral region, we used the well-established Castellvi classification to lumbosacral transitional vertebrae mainly based on the morphology of the transverse process of the last lumbar vertebra and whether it is fused with the sacrum or not.[31] Owing to the Castellvi classification we did not segment vertebrae that showed partial fusion with the sacrum (Castellvi grades III and IV) and did not include them for further analysis. Of note, in our database search, no "lumbalized" sacral vertebra with four sacral vertebrae remaining fused could be identified.

If present in the scan, vertebrae were labelled starting at the first cervical vertebra. If T1 was not visible within the scan's field-of-view, the thoracic spine was considered to have 12 vertebrae, as enumeration errors in the thoracic spine are much less frequent compared to the lumbar spine. [28]

# Data Records

## *Data Repositories and Storage*

The Verse 2020 dataset comprises 319 CT image series from 300 subjects and 4142 vertebrae encompassing 581 cervical, 2255 thoracic, and 1306 lumbar vertebrae as listed in Table 1. The stratification of anatomical variants in the VerSe dataset along with the corresponding ratings is listed in Table 2. The Dataset with its division into test, training, and private data subsets has been made publicly available under the creative commons license CC BY-SA 4.0 hosted at the open science framework https://osf.io/t98fz/. More information regarding the segmentation challenge algorithms submitted by the participants of the MICCAI VerSe challenges in 2019 and its second iteration in 2020 can be found at https://verse2019.grand-challenge.org/ and https://verse2020.grand-challenge.org/ as well as in the publication by Sekuboyina et al..[19] Worth of note, there is an overlap of 86 subjects and 105 imaging series between the VerSe 2020 dataset and the previously published VerSe 2019 imaging dataset, which is separately available for public use under the creative commons license CC BY-SA 4.0 at https://osf.io/nqjyw/. In a previous publication Loeffler et al. used the VerSe 2019 dataset to automatically detect and grade vertebral fractures and to calculate bone mineral density from a subset of the provided scans.[20]

## *Data Structure and File Formats*

All medical imaging files were converted into Neuroimaging Informatics Technology Initiative (NIfTI) format (https://nifti.nimh.nih.gov/nifti-1). Segmentation masks are also saved in NIfTI format and labels of all 4142 segmented vertebrae are provided in JSON format.

For organizational reasons of the segmentation challenge, all CT data (NIfTI format) in the Verse dataset was separated into a training dataset (100 subjects), a public test dataset (100 subjects), and a private test dataset (100 subjects) as previously described and demonstrated in Figure 1. Corresponding metadata is provided in the additional documents with the datasets.

## CT Imaging and Scan Provenience

CT scans included were intentionally chosen to be heterogeneous to ensure the best possible training and generalization of the algorithms. Therefore, data from the four major scanner vendors including Philips, Siemens, Toshiba, and GE from a variety of different multidetector CT scanner types of each vendor was included. The majority of images (45.8%) was acquired by Philips, 32.3% by Siemens, 6.3% by Toshiba, and 6.3% by GE scanners as shown in Table 1 and, on a patient-level in Supplement 1. There is no information regarding the scanner vendor provenience of the Glocker dataset[23], therefore they are listed as scans of "unknown" origin. Part of these examinations was carried out with additional administration of oral and/or intravenous contrast medium of various manufacturers. All included imaging series were based on edge-enhancing reconstructions, as this is the clinical standard for bone CT-imaging. Because the dataset comprises multi-centre imaging data from different scanner vendors, isotropic data was not available in all cases, replicating a typical clinical scenario.

## Anatomical Variants

From the overall 300 subjects with 319 scans, 90 scans (28.2%) demonstrated stump ribs. 135 scans visualized at least the lower part of the cervical spine and of these, 18 scans (13.3%) showed cervical ribs. From a total of 145 scans showing the complete thoracic spine, 131 scans presented with 12 thoracic vertebrae (90.3%), 8 scans showed 11 thoracic vertebrae (5.5%), and 6 scans demonstrated with 13 thoracic vertebrae (4.1%). From the 252 scans depicting the lumbosacral region, 165 showed 5 lumbar vertebrae (65.5%), 3 scans presented with 4 lumbar vertebrae (1.2%), and 85 scans showed 6 lumbar vertebrae (33.7%). Lumbosacral transitional vertebrae were graded according to the Castellvi classification, detailed in Table 2. In accordance with other authors and due to the difficulty of classifying Castellvi type I as well as the lack of clinical relevance, only Castellvi II – IV were classified.[32-34] Anatomical variants for each subject and scan are listed in Supplement 1.

# Technical Validation

The presented medical imaging data was derived from the institutional picture archiving system and therefore fully complies with the legal standards and quality controls for the acquisition of medical imaging in Germany and the European Union, as well as the industrial standards of the scanner vendors. Segmentation masks were prepared and annotated at voxel-level by a human-machine hybrid algorithm with manual checks and corrections by specifically trained medical students. Afterwards, the masks were reviewed, corrected, and finally approved by a neuroradiologist. The anatomical ratings were carried out by two neuroradiologists in a consensus reading. The resulting NIFTI datasets have successfully been processed by all dockers of the 13 participants of the VerSe 2020 challenge. Of these, the best performing docker achieved a mean vertebral identification rate of 95.6% with a mean localisation error of less than 2mm. Concerning segmentation, the best mean Dice score of 91.7%.

## Usage Notes

Additional metadata e.g., additional anatomical ratings, fracture grading, bone mineral density measurements will be continuously updated and published at the open science framework website https://osf.io/t98fz/.

## Code Availability

The in-house developed segmentation algorithm which was trained based on the VerSe dataset is publicly accessible at https://anduin.bonescreen.de. The entire dataset and its helper code (including the data reading, writing, and evaluation scripts) can be accessed here: https://github.com/anjany/verse.

For an overview of the various algorithms submitted as part of the segmentation challenges held at the MICCAI conferences in 2019 and 2020 based on the public VerSe datasets, kindly refer to Sekuboyina et al..[19] These nicely demonstrate the feasibility and potential of automated segmentation tasks and processing of spine CT imaging.

# Acknowledgements

This work was supported by the European Research Council (ERC) with Starting Grant No. 637164 "iBack" to Jan S. Kirschke and by the Deutsche Forschungsgemeinschaft (DFG, German Research Foundation) with Project No. 432290010 to Thomas Baum.

# Author contributions

| | |
|---|---|
| HL | manuscript preparation, data review/review, segmentation corrections, image rating |
| DS | manuscript preparation, data review |
| AS | code development, project design, data review, manuscript preparation |
| LM | data preparation/review, manuscript preparation, segmentation |
| ML | code development, project design, manuscript preparation, data review/preparation |
| AB | code development, manuscript preparation, data review/preparation |
| MH | code development, manuscript preparation, data review/preparation |
| GT | code development, project design, manuscript preparation, data review/preparation |
| KG | data preparation/review, manuscript preparation, segmentation |
| EN | data preparation/review, manuscript preparation, segmentation |
| TB | manuscript preparation, data review, algorithm/code development |
| CZ | data review, manuscript preparation, study design |
| JK | manuscript preparation, data review, segmentation corrections, image rating, algorithm/code development, study design |

## Competing interests

Hereby all authors state that there are no conflicts of interest regarding the presented manuscript and published data

## Tables and Figures

Table 1: Subject characteristics of the VerSe 2020 dataset and subset stratification. *Unknown scans were included from a public dataset[19].

Table 2: Subjects with cervical, thoracolumbar and lumbosacral anatomical variants. Lumbosacral vertebrae graded according to the Castellvi Classification.

Figure 1: Composition of the VerSe 2020 dataset: original data derived from the preexisting VerSe 2019 dataset, the Glocker[21] dataset and newly added subjects based on a pacs search for subjects with anatomical variants of the spine as well as images from different scanner vendors and imaging centres

Figure 2: Schematic overview on the image processing by the in-house developed algorithm publicly accessible under https://anduin.bonescreen.de (green boxes indicate fully automated algorithms) and the steps of manual interaction (blue boxes).

## Supplementary Material

Supplemental Table 1 lists the main characteristics of all VerSe (2019 and 2020) datasets on a patient level.

Figure 1

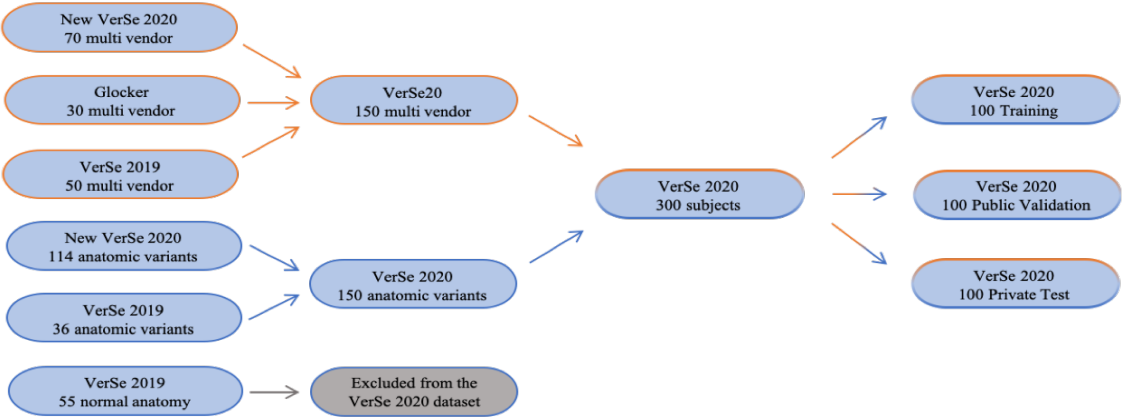

Figure 2

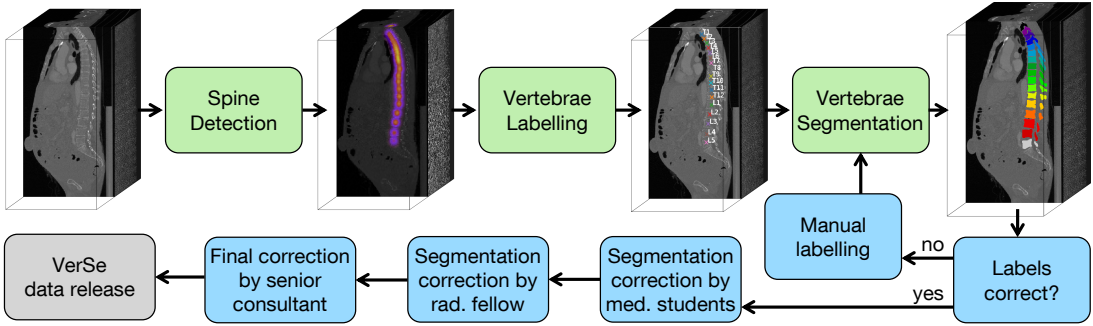

Table 1:

| VerSe 20 dataset | Private Test subset | Public Validation subset | Public Training subset | All |
|---|---|---|---|---|
| Number of patients (n) | 100 | 100 | 100 | 300 |
| Gender female/male (n) | 45/55 | 49/51 | 50/50 | 144/156 |
| Mean age (n ± SD) | 57.9 ± 17.6 | 54.5 ± 17.2 | 56.3 ± 18 | 56.2 ± 17.6 |
| Number of image series (n) | 103 | 103 | 113 | 319 |
| labeled vertebrae (n): | 1348 | 1366 | 1428 | 4142 |
| cervical vertebrae (n) | 193 | 164 | 224 | 581 |
| thoracic vertebrae (n) | 728 | 770 | 757 | 2255 |
| lumbar vertebrae (n) | 427 | 432 | 447 | 1306 |
| Image series from different vendors: | | | | |
| In-house   Philips (n) | 40 | 49 | 57 | 146 |
| In-house   Siemens (n) | 23 | 14 | 36 | 73 |
| External   GE (n) | 10 | 10 | 0 | 20 |
| External   Siemens (n) | 10 | 10 | 10 | 30 |
| External   Toshiba (n) | 10 | 10 | 0 | 20 |
| External   Unknown* (n) | 10 | 10 | 10 | 30 |

Table 2:

|  | Private Test subset | Public Validation subset | Public Training subset | All |
|---|---|---|---|---|
| Image series with cervical ribs (n) | 5 | 7 | 6 | 18 |
|   1 cervical rib | 3 | 5 | 4 | 12 |
|   2 cervical ribs | 2 | 2 | 2 | 6 |
|  |  |  |  |  |
| Thoracolumbar variants |  |  |  |  |
|   12 thoracic vertebrae (n) | 44 | 44 | 43 | 131 |
|   11 thoracic vertebrae (n) | 1 | 4 | 3 | 8 |
|   13 thoracic vertebrae (n) | 2 | 2 | 2 | 6 |
|  |  |  |  |  |
| Lumbosacral variants |  |  |  |  |
|   5 lumbar vertebrae (n) | 52 | 54 | 59 | 165 |
|   4 lumbar vertebrae (n) | 1 | 2 | 0 | 3 |
|   6 lumbar vertebrae (n) | 29 | 38 | 28 | 85 |
|  |  |  |  |  |
| Image series with stump rib (n) | 28 | 28 | 34 | 90 |
|   stump rib T12 (n) | 8 | 12 | 17 | 37 |
|   stump rib T13 (n) | 2 | 2 | 1 | 5 |
|   stump rib L1 (n) | 18 | 14 | 16 | 48 |
|  |  |  |  |  |
| Grading of the lumbosacral region |  |  |  |  |
|   Castellvi 0/1 (n) | 46 | 52 | 48 | 146 |
|   Castellvi 2a (n) | 10 | 8 | 17 | 35 |
|   Castellvi 2b (n) | 10 | 8 | 7 | 25 |
|   Castellvi 3a (n) | 5 | 7 | 2 | 14 |
|   Castellvi 3b (n) | 8 | 8 | 12 | 28 |
|   Castellvi 4 (n) | 3 | 2 | 2 | 7 |